\documentclass[prl,twocolumn,shopacs]{revtex4}%
\usepackage{amsmath}
\usepackage{amsfonts}
\usepackage{amssymb}
\usepackage{graphicx}%
\begin{document}
\title{Nonlinear screening theory of the Coulomb glass}
\author{Sergey Pankov}
\affiliation{Laboratoire de Physique Th{\'e}orique, Ecole Normale Sup{\'e}rieure, 24 Rue
Lhomond, 75231 Paris Cedex 05, France}
\author{Vladimir Dobrosavljevi\'c}
\affiliation{Department of Physics and National High Magnetic Field Laboratory, Florida
State University, Tallahassee, FL 32306}
\date{\today{}}


\begin{abstract}
A nonlinear screening theory is formulated to study the problem of gap
formation and its relation to glassy freezing in classical Coulomb glasses. We
find that a pseudo-gap ("plasma dip") in a single-particle density of states
begins to open already at temperatures comparable to the Coulomb energy. This
phenomenon is shown to reflect the emergence of short range correlations in a
liquid (plasma) phase, a process which occurs even in the absence of disorder.
Glassy ordering emerges when disorder is present, but this occurs only at
temperatures roughly an order of magnitude lower. Our result demonstrate that
the formation of the "plasma dip" at high temperatures is a process distinct
from the formation of the Efros-Shklovskii (ES) pseudo-gap, which in our model
emerges only within the glassy phase.

\end{abstract}
\maketitle

The interplay of interactions and disorder remains one of the most important
open problems in condensed matter physics. These effects are most dramatic in
disordered insulators, where the pioneering work of Efros and Shklovskii (ES)
\cite{efros75} emphasized the fundamental significance of the long-ranged
nature of Coulomb interactions. This work presented convincing evidence that
at $T=0$ a soft "Coulomb gap" emerges in the single-particle density of states
(DOS) which, in arbitrary dimension $d$, reads 

\begin{equation}
g(\varepsilon)\sim\varepsilon^{d-1}.
\end{equation}

>From a general point of view this result is quite surprising. It
indicates a power-law distribution of excitation energies, i.e.
the absence of a characteristic energy scale for excitations above
the ground state. Such behavior is common in models with broken
continuous symmetry, where it reflects the corresponding Goldstone
modes, but is generally not expected in discrete symmetry models,
such as the one used by ES. Here, it may reflect unusually strong
frustration behavior inherent to Coulomb interactions in presence
of disorder.

Indeed, the ES model seems to display several glassy features characterized by
a large number of meta-stable states and slow relaxation, as clearly seen in
many simulations \cite{davies82,mobius92,grannan93,grempel}, and even in some
experiments \cite{glassexp}. Interestingly, a precursor of the gap begins to
appear \cite{grannan93,sarvestani95} already at relatively high temperatures,
while glassy ordering emerges only at much lower $T$. Similar behavior has
been identified even in absence of randomness \cite{efros92}. Is the physics
of the Coulomb gap thus related or is it unrelated to the glassy features of
the system? The close connection between the two phenomena was recently
demonstrated \cite{pastor99} for a mean-field model of interacting disordered
electrons in the limit of large coordination, but the issue remains unresolved
for Coulomb systems in finite dimensions where the ES theory applies.

To address these important issues in a controlled and precise
fashion, we make the following observation which is the main
physical point of this letter. We stress that the principal ES
result - the emergence of a power-law spectrum - is not specific
to low dimensions! Its physical origin and its relation to high
temperature anomalies can, therefore, be investigated by the 
theoretical approaches 
controlled in the limit of high
spatial dimensions. Such a theory for the Coulomb glass (CG) model is
presented in this letter. Our main conclusions are as follows: (1)
A non-universal pseudo-gap in the DOS (we call it the "plasma
dip") begins to emerge at temperatures of the order of the Coulomb
energy. It reflects strong short range correlations in the Coulomb
plasma, a feature that is most pronounced in absence of disorder
\cite{efros92}, but is unrelated to the Coulomb gap of ES. We
obtain simple analytical results that in quantitative detail
describe the temperature evolution of the plasma dip, in excellent
agreement with all existing simulations. (2) The high temperature
plasma (fluid) phase becomes unstable to ergodicity breaking at
temperatures typically ten times lower, as the system enters a
glassy state. We argue that a true ES pseudo-gap emerges only
within the glassy (nonergodic) phase, and that its scale-invariant
form reflects the marginal stability of such a glassy state.

\emph{Nonlinear screening theory.} The simplest many-body approach to Coulomb
systems is the well-known Debye-Huckel theory , which provides a linear
screening description equivalent to a Gaussian approximation for the plasmon
mode. This formulation, however, fails badly at low temperatures, where
nonlinear effects lead to strong correlations in the plasma phase. Even worse,
such a Gaussian theory is unable to describe glassy freezing even in the
well-understood mean-field limit corresponding to infinite range interactions.
To overcome these difficulties, we use the simplest theory of nonlinear
screening given by the classical limit of the so-called extended dynamical
mean-field theory \cite{edmft} 
, which also describes the leading order
nontrivial correlations in the limit of large coordination. In this approach,
the environment of a given site is approximated by free collective modes
(plasmons in our case), the dispersion of which is self-consistently
determined. In recent work, a version of this method has been successfully
applied to the problem of self-generated glassiness \cite{lopatin02} in
systems with frustrated phase separation \cite{schmalian} without disorder.
Here, we apply it to the classical Coulomb glass 
model \cite{efros75}
given by the Hamiltonian \
\begin{equation}
H=\sum_{i}\phi_{i}n_{i}+\frac{1}{2}\sum_{ij}V_{ij}(n_{i}-K)(n_{j}%
-K),\label{cghamiltonian}%
\end{equation}
where $n_{i}=0,1$ is the electron occupation number, and $\phi_{i}$ is a
Gaussian distributed random potential of variance $W^{2}$. We express the
Coulomb interaction $V_{ij}={\varepsilon_{0}}/{r_{ij}}$ in units of the
nearest-neighbor repulsion ${\varepsilon_{0}}$, and the inter-site distance
$r_{ij}$ in units of the lattice spacing. We adopt the following notation
throughout the paper. For vectors and matrices in the replica space we use
bold font and the hat symbol correspondingly. In the replica symmetric ansatz
(RS) a matrix $\hat{O}\equiv\{O_{c},O\}$ is parameterized by its connected
part $O_{c}$ and its off-diagonal part $O$. Thus for the density density
correlator we use $\hat{q}\equiv\{\chi,q\}$. Thermal and disorder averages are
denoted as $\langle O\rangle_{T}$ and $[O]_{dis}$ respectively, and $\langle
O\rangle\equiv\lbrack\langle O\rangle_{T}]_{dis}$.

To derive the desired self-consistency equations we 
average over 
disorder using the standard replica method \cite{pastor99}, and use a cavity
construction \cite{pastor99,edmft}, integrating out the degrees of freedom on
all sites except the considered one. The resulting contribution to the local
effective action is computed in the Gaussian approximation, giving a term of
the form $-\frac{1}{2}\delta\mathbf{n}\hat{\Delta}\delta\mathbf{n}$, where
$\delta\mathbf{n}=\mathbf{n}-\langle\mathbf{n}\rangle$. By requiring that the
local density-density correlator $q^{\alpha\beta}=\langle\delta n^{\alpha
}\delta n^{\beta}\rangle_{c}$ is correctly reproduced by the effective action,
one obtains a self-consistency condition. We get
\begin{align}
&  q^{\alpha\beta}=\langle\delta n^{\alpha}\delta n^{\beta}\rangle_{cS_{eff}%
},\nonumber\\
&  S_{eff}=-\frac{1}{2}\delta\mathbf{n}\hat{\tilde{\Delta}}\delta
\mathbf{n},\nonumber\\
&  \hat{q}=\sum_{k}\left(  {\hat{q}}^{-1}+\hat{\Delta}-\beta V_{k}\right)
^{-1},\label{edmfteqs}%
\end{align}
where ${\tilde{\Delta}}^{\alpha\beta}={\Delta}^{\alpha\beta}+\beta^{2}W^{2}$,
and $V_{k}$ is the Fourier transform of the interaction potential. In case of
the RS solution $\hat{\tilde{\Delta}}\equiv\{\Delta_{c},\beta^{2}W_{eff}%
^{2}\}$, where $W_{eff}=\sqrt{W^{2}+\beta^{-2}\Delta}$ is the renormalized disorder.

\emph{Fluid solution.} To examine the evolution of the system in the high
temperature phase we first examine the RS solution. In the $n\rightarrow0$
replica limit we get
\begin{align}
&  q=\frac{1}{4}\int D[x]\tanh^{2}\left(  \frac{1}{2}x\beta W_{eff}\right)
,\label{qeq}%
\end{align}
where 
$D[x]\equiv(2\pi)^{-1/2}\exp
\{-x^{2}/2\}\;dx$. The self-consistency condition becomes
\begin{align}
&  \chi+q=\frac{1}{4},\nonumber\\
&  \chi=\sum_{k}\left(  {\chi}^{-1}+\Delta_{c}-\beta V_{k}\right)
^{-1},\nonumber\\
&  q=\left(  q\chi^{-2}-\Delta\right)  \sum_{k}\left(  {\chi}^{-1}+\Delta
_{c}-\beta V_{k}\right)  ^{-2},\label{sceq}%
\end{align}

The Eqs. (\ref{qeq},\ref{sceq}) can easily be solved numerically 
for the parameters $\chi,q,\Delta_{c},\Delta,W_{eff}$, 
to calculate the density of states (DOS)
function $g(\epsilon)$, to examine the stability of the RS solution, and to
compute the entropy $S$. The above equations are written for the half filled
case in absence of charge ordering. Generalization to uniformly ordered
phases is a standard procedure \cite{chitra01pankov02}, where the ordering
transition is signaled by a divergence in $\chi_{k}=\left(  {\chi}^{-1}%
+\Delta_{c}-\beta V_{k}\right)  ^{-1}$ at the ordering vector $k=Q$. For
simplicity, most of our results are written in the homogeneous phase at half
filling and that will be assumed unless stated otherwise.

\emph{Density of states.} For the CG model, the single particle
DOS (tunneling DOS) function $g(\epsilon)$ is simply given by the
distribution of the local fields (energies) $\epsilon_{i}=\partial
H/\partial n_{i}$:
\begin{equation}
g(\epsilon)=\sum_{i}\langle\delta(\epsilon-\epsilon_{i})\rangle.\label{dos}%
\end{equation}
Integrating out all sites except one, we derive an expression for
$g(\epsilon)$ in terms of the local effective field $\hat{\tilde{\Delta}}$.
The final result for the RS solution reads:
\begin{multline}
g(\epsilon)=\int D[x]\frac{\beta}{\sqrt{2\pi\Delta_{c}}}\,\frac{\cosh{\frac
{1}{2}\beta\epsilon}}{\cosh{\frac{1}{2}x\beta W_{eff}}}\\
\times\exp{\left\{  -\frac{1}{2\Delta_{c}}\left[  \frac{1}{4}\Delta_{c}%
^{2}+\left(  \beta\epsilon+x\beta W_{eff}\right)  ^{2}\right]  \right\}
.}\label{dosformula}%
\end{multline}
In the cavity method langauge\cite{mezard86} this result can be
interpreted as the Gaussian distribution of the cavity fields
(different from the local fields) of variance $W_{eff}$, modified
by the the (Onsager) self reaction term representing the plasma
correlations.

We get insight into the behavior of the DOS by considering some analytically
solvable limits. At $T\rightarrow0$, the DOS remains finite at the Fermi
level, though it can be exponentially small for weak disorder. This shows that
the fluid (RS) solution does not capture the physics of the true ES gap, which
emerges only within the glassy phase. On the other hand, as long as the RS
solution is stable, a large "plasma dip" may develop, but it will have no
direct relation to the glassy physics or the ES gap. It reflects strong
short-range correlations in the Coulomb plasma (fluid) phase, which are
suppressed at large disorder. Here, the RS DOS reduces to the bare disorder
distribution. In the opposite limit of vanishing disorder $W_{eff}%
\rightarrow0$ the DOS expression simplifies and for an arbitrary
filling reads:
\begin{multline}
g(\epsilon)= \frac{\beta}{\sqrt{2\pi\Delta_{c}}} \exp\left(
-\frac{\frac{1}{4}\Delta
_{c}^{2}+\beta^{2}\epsilon^{2}}{2{\Delta_{c}}}\right) \\
\times\left[  \cosh{\frac{1}{2}\beta\epsilon}-
(2K-1)\sinh{\frac{1}{2}\beta\epsilon}\right].
\label{dosW0}%
\end{multline}

\begin{figure}[ptb]
\includegraphics[width=3.3in,height=4in]{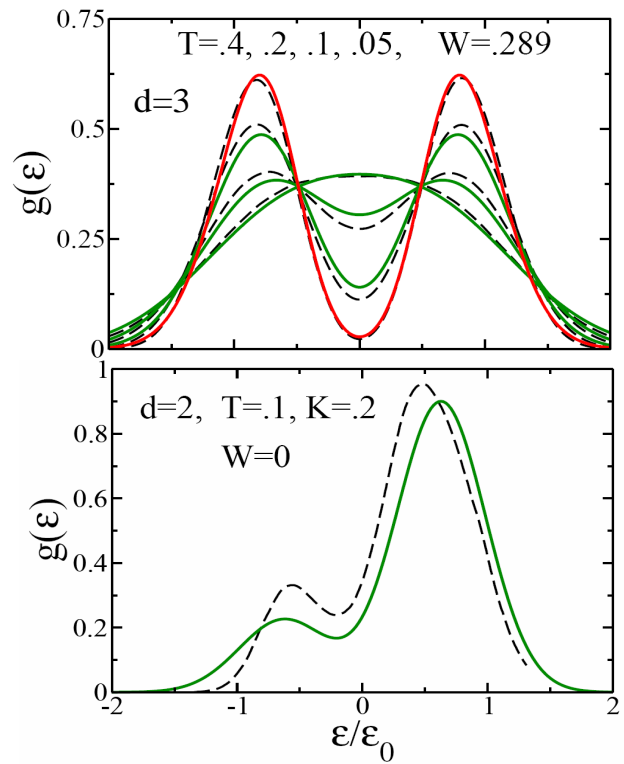} \caption{ Our analytical
predictions for the single-particle density of states (full lines) are found
to be in excellent quantitative agreement with simulation results (dashed
lines), with no adjustable parameters. Shown are results for the three
dimensional case studied in from Ref. \cite{sarvestani95}, corresponding to
$W=1/(2\sqrt{3})$, and temperatures $T=0.4,0.2,0.1,0.05$ (top panel), and the
two dimensional model of Ref. \cite{efros92}, corresponding to $W=0$, $T=0.1$,
$K=0.2$. All lines correspond to the plasma phase, while $T=0.05$ is
very close to the glass transition temperature. }%
\label{compare_vojta}%
\end{figure}

To support the validity of our formulation, we compare our analytical results
with available numerical simulations. In Fig. \ref{compare_vojta} (top
panel), we examine the situation studied in Ref.\cite{sarvestani95}, where
calculations were done for a 3D CG on a cubic lattice, for a set of
temperatures in the fluid phase. The lowest temperature is very near 
the glass transition temperature (see our phase diagram, Fig. 2). 
As we can see, our theory captures in surprisingly quantitative 
detail the formation of the "plasma dip" in the fluid phase. 
In the past, this phenomenon has often been confused with the
formation of the true ES gap which, as we argue below, only emerges within the
glassy phase. Finally, we test limits of our theory by computing the DOS for
the 2D CG in absence of disorder. Even in this extreme case, we
reproduce semi-quantitatively exact numerical results of Ref.
\cite{efros92}. \vspace{0.5cm}

\begin{figure}[t]
\includegraphics[width=3.3in,height=2.4in]{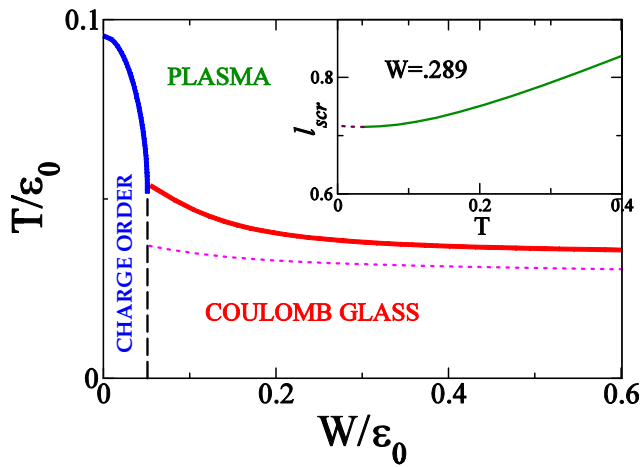} \caption{3D Coulomb glass
phase diagram. The full horizontal line indicates the RSB instability and the
dotted line shows where the RS entropy turns negative. The screening length in
the inset is plotted for the same disorder and the range of temperatures
(fluid phase), as in Fig.\ref{compare_vojta}.}%
\label{phasediagram}%
\end{figure}

\emph{Glassy ordering.} To examine the stability of the fluid phase to glassy
ordering, we examine the Baym Kadanoff (BK) functional $\Gamma_{BK}[\hat{q}]$.
This is a functional of the correlator $\hat{q}$, which yields the exact
equations of motion at the saddle point, where it coincides with the exact
free energy. To obtain our self-consistency conditions, a local approximation
\cite{chitra01pankov02} is made on the two particle irreducible part of
$\Gamma_{BK}$. In this formulation, the stability of our fluid RS solution can
be obtained by a standard replica symmetry breaking (RSB) analysis
\cite{almeida78} of the BK functional at the saddle point. The corresponding
RSB instability criterion takes the form
\begin{equation}
\frac{1}{\left[  \chi_{ii}^{2}\right]  _{dis}}-\frac{1}{[\chi_{ii}]_{dis}^{2}%
}+\frac{1}{\sum_{j}[\chi_{ij}]_{dis}^{2}}=0.\label{rsbcondition}%
\end{equation}
Here, $\chi_{ij}$ is the density-density correlation function computed for a
fixed realization of disorder, i.e. $\chi_{ij}=\langle\delta n_{i}\delta
n_{j}\rangle_{T}-\langle\delta n_{i}\rangle_{T}\langle\delta n_{j}\rangle_{T}%
$. The left hand side of Eq. (\ref{rsbcondition}) is nothing but $1/{\sum
_{j}[\chi_{ij}^{2}]_{dis}}$, the inverse of the glass susceptibility, a
quantity which diverges at the transition. In terms of the RS solution the RSB
condition reads
\begin{equation}
\frac{q}{\Delta}=\frac{1}{16}\int D[x]\cosh^{-4}\left(  \frac{1}{2}x\beta
W_{eff}\right)  .\label{rsrsbcondition}%
\end{equation}
As an illustration, we present results for the CG on a 3D cubic
lattice, and in Fig. \ref{phasediagram} we plot the corresponding phase
diagrams obtained by numerically solving our self-consistency
conditions. 
At small disorder and temperature $T\approx.95$ (which is in satisfactory
agreement with the exact value\cite{mobius03} $T_c=.129$) the system enters 
the charge ordered phase.
phase. Stronger disorder suppresses the charge ordering, and the system can
exist either in a liquid phase (at higher temperature) or in the glass phase
(at lower $T$). The liquid is separated from the glass by the RSB line, also
known as the Almeida-Thouless \cite{almeida78} line. We emphasize that the
ordering temperature we predict is roughly an order of magnitude smaller then
the Coulomb energy, in remarkable quantitative agreement with all available
simulation results \cite{mobius92,efros92,grannan93,grempel}.

This interesting fact can be traced down to the screening of the Coulomb
interaction. Indeed, the overall energy scale characterizing the screened
Coulomb potential $V_{scr}(r)= \varepsilon_{0} \exp\{-r/\ell_{scr} \}/r$ is
roughly an order of magnitude smaller then the bare Coulomb energy. The
corresponding screening length $\ell_{scr}= [( \chi^{-1}+\Delta_{c}%
)/(\beta\varepsilon_{0})]^{1/2}$ (shown for $W=(2\sqrt{3})^{-1}$ in the inset
of Fig.~2) decreases (albeit weakly) with temperature, and remains short
throughout the fluid (RS) phase. This observation also makes it clear why the
true ES gap cannot emerge in absence of glassy ordering. The screening
mechanism remains operative throughout the (ergodic) fluid phase, and thus the
long-range character of the Coulomb interaction, which is crucial for the ES
argument remains inoperative. In contrast, within the glass phase, following
the arguments from Ref. \cite{pastor99}, we expect the relevant zero-field
cooled compressibility to decrease and vanish at $T=0$. This mechanism opens a
route for the screening to be suppressed at low temperatures, and the true ES
gap to emerge.

To provide further evidence of the instability of the fluid phase to glassy
ordering, we also calculate the entropy in the fluid (RS) solution, which
takes the form
\begin{multline}
S=\int D[x]\ln{\left[  2\cosh{\left(  \frac{1}{2}x\beta W_{eff}\right)
}\right]  }\\
+\frac{1}{2}\sum_{k}\ln{\chi_{k}}-\frac{1}{2}\ln{\chi}-2\chi\beta
W_{eff}.\label{entropy}%
\end{multline}
It is well known that for standard mean-field glass models, the RS replica
theory predicts negative entropy at $T=0$, while the lower bound for the glass
transition temperature can be set where the RS entropy changes sign. It is
easy to show from Eq. (\ref{entropy}) that our RS entropy proves strictly
negative at $T=0$ as well. In Fig. 2 we also plot (dashed line) a lower bound
for the RSB instability line where the RS entropy turns negative, providing
further evidence that the fluid phase cannot survive down to $T=0$.

\emph{The glass phase and the Efros-Shklovskii gap.} In this letter we do not
explicitly examine the RSB solution of our model. Nevertheless, we follow
arguments similar to those of Ref. \cite{pastor99}, and use the large disorder
asymptotics of the glass transition line to determine the powerlaw form of the
$T=0$ Coulomb gap in the glassy phase. At $W\gg\varepsilon_{0}$, we find
\begin{equation}
T_{G}\sim\varepsilon_{0}^{1+\frac{1}{\gamma}}W^{-\frac{1}{\gamma}%
},\label{lrATasympt}%
\end{equation}
where the exponent $\gamma$ is \emph{identical} to that predicted by ES for
the $T=0$ DOS in the CG
\begin{equation}
g(\epsilon)\sim\varepsilon_{0}^{-1-\gamma}\epsilon^{\gamma}.\label{esbound}%
\end{equation}
For an interaction of a general power law form $V(r)\sim\varepsilon_{0}/r^{a}%
$, the ES argument predicts $\gamma=(d-a)/a$ in arbitrary dimension $d$. The
asymptotic regime, however, sets in at larger values of disorder, not shown in
the Fig.(\ref{phasediagram}).

Let us explain the importance of Eq. (\ref{lrATasympt}). We have seen that for
$W\neq0$, the DOS remains finite in the RS fluid phase, so a true ES
pseudo-gap can emerge only due to glassy ordering. For another electron glass
model, work of Ref. \cite{pastor99} has established that the emergence of a
true pseudo-gap at $T=0$ directly follows from the marginal stability of the
glassy state, and that its form also determines the large disorder asymptotics
of $T_{G} (W)$. Given the close similarity of our mean-field equations for the
CG model to those examined in Ref. \cite{pastor99}, we expect the same
mechanism to apply here as well. Using the expected form of the ES gap, we can
estimate the glass transition temperature as the energy scale $E_{gap}$
characterizing the "width" of the gap that opens in the low temperature phase.
We find $E_{gap}\sim\varepsilon_{0}^{1+\frac{1}{\gamma}} W^{-\frac{1}{\gamma}%
}$, coinciding with Eq. (\ref{lrATasympt}). This result presents strong
evidence in favor of the close relation between glassy ordering and the
emergence of the ES gap.

In conclusion, we have formulated a simple many-body theory that is able to
clarify the relation between the finite temperature formation of the Coulomb
gap and the emergence of glassy ordering in disordered Coulomb systems in
finite dimensions. This nonlinear screening approach is flexible enough to
allow for future extensions to quantum models \cite{pastor99,vojta}, and to
study the role of Anderson and Mott localization \cite{mitglass} in Coulomb systems.

\begin{acknowledgments}
We thank G. Biroli, A. Georges, D. Grempel, M. Muller, D. Popovi\'{c}, B.
Shklovskii, T. Vojta, and G. Zimanyi for useful discussions. This work at FSU
was supported by the NSF grant DMR-0234215. SP acknowledges support from the
CNRS, France. After this work has been completed, the authors became aware of
the closely related calculation of M. Muller and L. B. Ioffe
(cond-mat/0406324), where complementary results consistent with our findings
were presented.
\end{acknowledgments}

\end{document}